\documentclass[prl,floatfix,twocolumn,superscriptaddress]{revtex4-1}
\usepackage{graphicx}
\usepackage[rgb,dvipsnames]{xcolor}
\usepackage{amsmath, amsthm, amssymb}
\usepackage{epsfig}
\usepackage{float}
\usepackage{hyperref}
\usepackage[normalem]{ulem}

\usepackage{placeins}
\usepackage[symbol]{footmisc}

\usepackage{tikz}
\usetikzlibrary{trees}

\begin{document}

\title{Disorder in cellular packing can alter proliferation dynamics to regulate growth}

\author{Chandrashekar Kuyyamudi}
\affiliation{The Institute of Mathematical Sciences, CIT Campus, Taramani, Chennai 600113, India}
\affiliation{Homi Bhabha National Institute, Anushaktinagar, Mumbai 400094, India}
\author{Shakti N. Menon}
\affiliation{The Institute of Mathematical Sciences, CIT Campus, Taramani, Chennai 600113, India}
\author{Fernando Casares}
\affiliation{CABD, CSIC-Universidad Pablo de Olavide-JA, 41013 Seville, Spain}
\author{Sitabhra Sinha}
\affiliation{The Institute of Mathematical Sciences, CIT Campus, Taramani, Chennai 600113, India}
\affiliation{Homi Bhabha National Institute, Anushaktinagar, Mumbai 400094, India}
\date{\today}

\begin{abstract}
Controlling growth via cell division is crucial in the development of higher organisms, and yet the mechanisms through which this is achieved, e.g., in epithelial tissue, is not yet fully understood. We show that by coupling the cell cycle oscillator governing cell division to signals that encode inter-cellular contacts, this phenomenon can be seen as a collective dynamical transition in a system of coupled oscillators in lattices with changing degree of disorder. As the distribution of cellular morphological characteristics become more homogeneous over the course of development, the contact-induced signals to the cells increase beyond a critical value to trigger coordinated cessation of oscillations, eventually leading to growth arrest.  Our results suggest that the global phenomenon of growth rate reduction as a tissue approaches its appropriate size is causally related to the increasingly regular geometry of local cell-cell contact interfaces.
\end{abstract}
\maketitle

\begin{figure}[tbp!]
\includegraphics{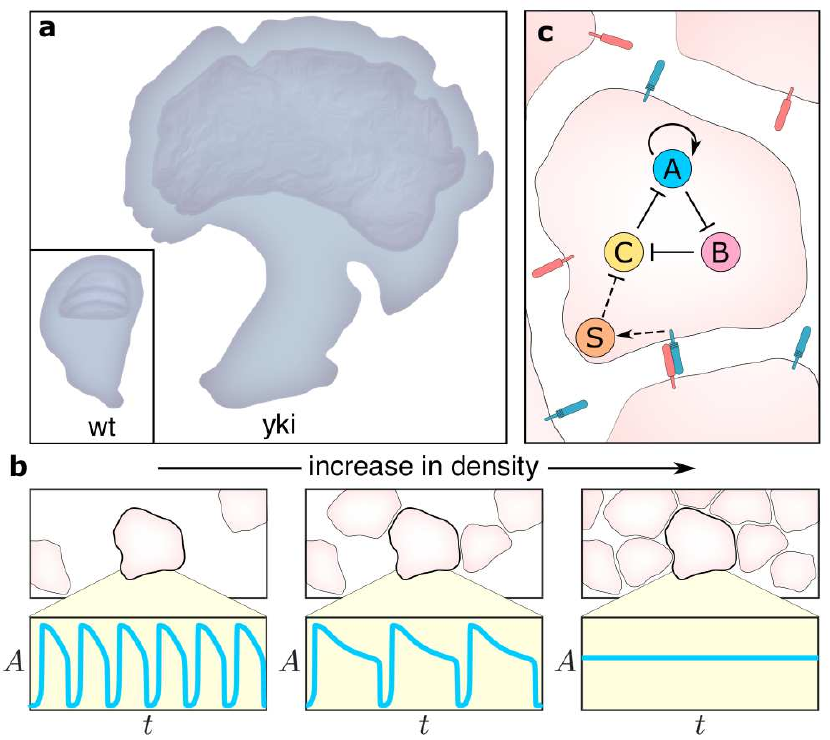}
\caption{\textbf{Increased contact with neighbors over time results in a decreasing rate of cell division, culminating in growth arrest.}
(a) Schematic diagram showing an overgrown \textit{Drosophila} imaginal disc
(compared to the wild-type, wt, shown in inset) resulting from the
over-expression of \textit{yorkie} (yki), the main transcriptional effector of the
Hippo signaling pathway (figure adapted from Ref.~\cite{Huang2005}).
(b) The time-varying concentration $A$ of a representative molecular species
constituting the
oscillator of a cell (indicated by the bold outline) is shown at three instances
in which progressively larger proportions of its surface are in contact with adjacent cells.
The frequency of the oscillations (governing the rate of cell division) decreases as the total
area in contact with neighbors increases, eventually culminating in oscillator death,
and hence termination of further cell division.
(c) Proposed mechanism for the regulation of cell cycle oscillations by
contact-mediated signaling.
The trans-membrane receptors and ligands are represented by blue and orange rods, respectively.
A receptor binding to a ligand from a neighboring cell triggers a signaling cascade
whose terminal effector molecule S regulates the activity of the cell cycle oscillator,
represented by the loop comprising molecules A, B and C.
}
\label{figure1}
\end{figure}

Growth, or the gradual increase in size,
is a fundamental attribute of all living systems~\cite{Gilbert2018}. A central question in developmental biology is how does the body and its parts
know that they have attained their appropriate size so that further growth can stop~\cite{Wolpert2011,Shahbazi2020}.
The problem is essentially analogous to that of quorum sensing~\cite{DeMonte2007,Taylor2009,Whiteley2017}: how do cells, which have access only to local information from their neighborhood, respond to changes in global properties 
of the collective they are part of~\cite{Lander2013,Kuyyamudi2021}.
As tissues and organs are composed of large numbers of cells, each implementing
intrinsic programs to regulate their division, arresting growth will require coordination achieved through self-organization
across large assemblages~\cite{Hagan2021}. 
Failure to achieve this can have consequences not only during
development of an organism, e.g., resulting in potentially fatal deformities
[Fig.~\ref{figure1}~(a)] but also later
in the adult stage, when unchecked growth in the number of cells often leads to cancer~\cite{Sherr1996,Kuznetsov2000}.
As cell proliferation through mitotic cell division is the primary means
by which growth occurs~\cite{Wolpert2011}, it appears that controlling the cell cycle which regulates mitosis is
the key to ensuring the appropriate final size for any
developing system.
As the transitions between different stages of the cell cycle are governed by oscillations in
the concentrations of proteins known as cyclins~\cite{Santamaria2007,Orlando2008}, preventing further cell division once the system
has reached optimal size requires a mechanism by which coordinated cessation of these
intra-cellular oscillations can be achieved.
While it is known that increasing cell density can eventually arrest growth 
via contact inhibition of proliferation (CIP)~\cite{Stoker1967,Sestan1999,Zhao2007,Kim2011,Puliafito2012,Aoki2013}
[see Fig.~\ref{figure1}~(b)], in general the processes through which signals encoding 
inter-cellular contact events modulate the oscillatory dynamics are not fully understood.

Knowing when to stop growing becomes particularly challenging in tissues comprising epithelial cells,
which are present in most organs of the mammalian body~\cite{Wolpert2011}.
As adjacent cells remain in contact during the growth of epithelial sheets over the course of
development, contact inhibition cannot be invoked to explain the arrest of growth
when an organ approaches its final size~\cite{Honda2017}.
Understanding the process that stops further cell division in such systems is important as
uncontrolled proliferation in epithelia is linked to more than $85\%$ of all human cancers~\cite{Mccaffrey2011,Macara2014}.
Experiments have implicated the Hippo intra-cellular signaling pathway as a key coordinator
for the regulation of cell division, as it has been shown to inhibit the transcriptional co-activator protein \textit{yap}/\textit{yorkie} (in vertebrates/\textit{Drosophila}) resulting in suppression
of growth~\cite{Harvey2003,Huang2005,Thompson2006,Zhao2007,Zeng2008,Kango2009,Oh2010,Gumbiner2014}. Indeed, mutations resulting in loss of function of Hippo or over-expression
of \textit{yap}/\textit{yorkie} result in abnormal growth [Fig.~\ref{figure1}~(a)], which can
lead to cancer.
While it is generally believed that \textit{yap} and the Hippo pathway plays a role in transducing increased mechanical tension among cells that result from growth in order to inhibit proliferation~\cite{Ma2019}, the mechanism linking the two is yet to be fully understood.
As the morphological characteristics of cells (e.g., size and shape) continually change in a growing epithelial sheet, an intriguing possibility is that the
local geometry of cell-cell interfaces convey information about the state of the growing organ
to the intra-cellular signaling pathway which can eventually arrest the cell cycle
oscillations.
Thus, regulation of growth can be viewed as a collective dynamical transition in a system of coupled
oscillators forming a disordered lattice with a dynamically evolving contact geometry.

In this paper, we propose a unified framework that describes both CIP resulting from
increasing cellular contacts with rising density, as well as, growth termination triggered by
morphological changes in confluent epithelial sheets. We show that both phenomena
are consequences of arresting the cell cycle oscillator by signals resulting from cell-cell contact,
such as those mediated by the Hippo pathway. We show that the signal, whose intensity depends on the number of inter-cellular contacts (as per the model assumptions), can convey
information about the the geometry of the
interfaces. This allows changes in size and shape of cells in local neighborhoods to alter the
mean frequency of the cellular oscillators, and hence the overall growth rate of the
entire system.
We achieve this by modulating the activity of the cell cycle oscillator via
the concentration of an effector molecule, which encodes the
extent of inter-cellular contact. By deriving a closed-form expression for the oscillator
frequency as a function of the intensity of this signal and its strength of coupling to the oscillator,
we show that above a critical strength of the coupling, the system exhibits a transition
to oscillation arrest as the signal intensity increases.
We demonstrate that the termination of growth in epithelial sheets, whose cells can be modeled as polygons of different shapes and sizes, may come about through coordinated cessation of oscillations
as the contact-induced signals to individual cells alter in magnitude with changing
heterogeneity in cellular morphology over the course of
development.
The evolution of the distribution of the
morphological characteristics is implemented by generating
progressively more homogeneous space-filling arrangements of cells in lieu
of incorporating explicit cell division.
This allows
us to show that the progressive reduction of growth rate can be brought about exclusively by
changes in local cell-cell contact geometry that result from reduction in cellular heterogeneity
as the organ size increases. We also find that the strength of coupling defines two
contrasting regimes characterized by opposing responses of the growth rate to increasing
heterogeneity of cell shapes and sizes, a result that has intriguing implications for
pathologies, such as developmental dysplasia and cancer.

%
To explore how self-organized arrest of tissue growth can arise as a result of alterations
in local cellular density over the course of development, we investigate a model
of a two-dimensional sheet of cells, in which the cell cycle oscillations
governing proliferation, and hence growth
of tissue, are regulated by contact-mediated signaling [Fig.~\ref{figure1}~(c)].
When cells come in close physical proximity, receptors on the surface of a cell can bind
to membrane-bound ligands of a neighboring cell, eventually triggering a signaling cascade
(such as the Hippo pathway).
This is coupled to the cell cycle oscillator by assuming that a
downstream effector $S$ of the cascade, whose magnitude conveys information
about the local extra-cellular ligand concentration, represses one of the molecular components
of the oscillator.
For concreteness, we use a cell cycle oscillator model
involving three molecular species (one of which is considered to be self-activating) that
repress each other in a cyclic manner [Fig.~\ref{figure1}~(c)]. It is capable of
oscillating over a wide range of frequencies with an almost invariant amplitude~\cite{Tsai2008},
a desirable property in view of the fact
that cell division rates can vary widely within the same organism.
Expressing the concentrations of the activated forms of the molecules as
$A$, $B$ and $C$, the dynamics of the system upon coupling to $S$ (via $C$~\cite{note2})
can be described by the following set of equations:
\begin{eqnarray}
  \label{eq:A}
  \frac{dA}{dt} &=& k_1 (A_T - A) - \frac{k_2 C^h}{K^h + C^h} A \\\nonumber
  &+& k_7(A_T - A)\frac{A^h}{K^h + A^h} \,, \\
  \label{eq:B}
  \frac{dB}{dt} &=& k_3 (B_T - B) - \frac{k_4 A^h}{K^h + A^h} B \,, \\
  \label{eq:C}
  \frac{dC}{dt} &=& k_5 (C_T - C) - \frac{k_6 B^h}{K^h + B^h} C - \frac{k_8 S^g}{\Psi^g + S^g} C\,,
\end{eqnarray}
where $A_T,B_T,C_T$ correspond to the total concentration of the activated and inactivated
forms of the three molecular species, respectively. The parameters $k_1, \ldots, k_8$
represent rate constants, with $k_1, k_3, k_5$ and $k_7$ governing the transitions
to the activated forms of the molecules while $k_2$, $k_4$ and $k_6$ regulate
the inactivation processes mediated by the presence of the corresponding repressors.
The rate $k_8$ quantifies the coupling strength between the contact-induced signal $S$
and the cell cycle oscillator via the repression of the oscillator component C.
The inactivation of each molecular species by its repressor in the oscillator is modeled by a Hill
function, whose functional form is parameterized by the exponent $h$ and half-saturation
constant $K$. A similar Hill function is also used to regulate the inactivation of $C$ by $S$,
with the corresponding parameters being $g$ and $\Psi$.

\begin{figure}[htbp!]
\includegraphics{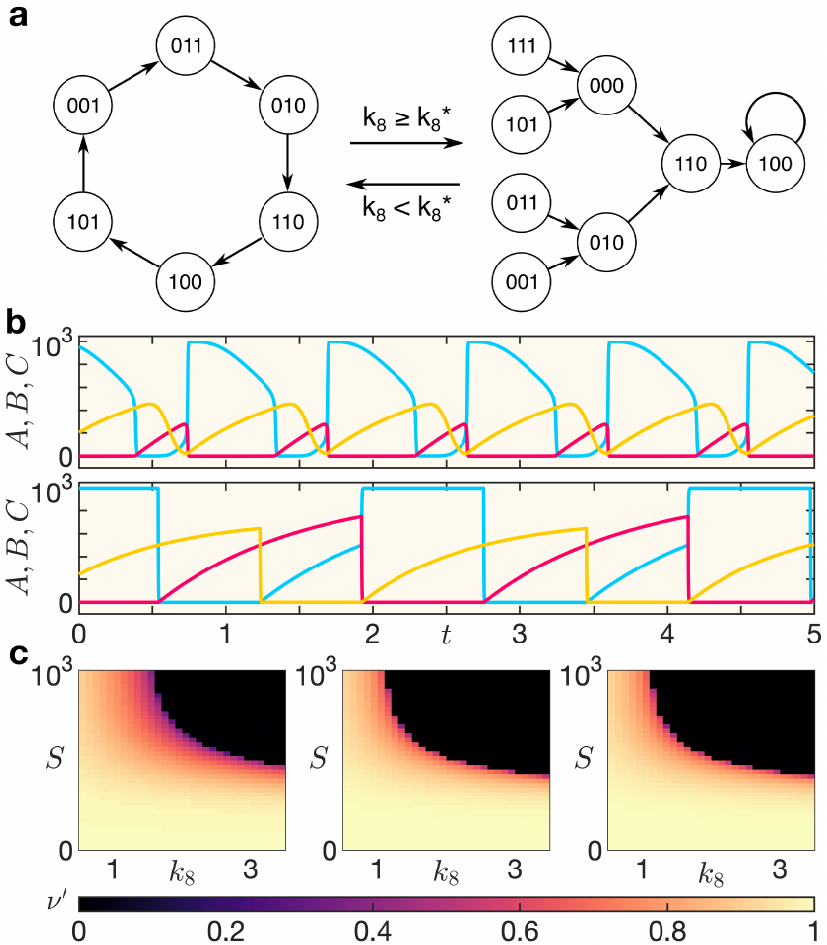}
\caption{\textbf{Coupling mediated dynamical transition from cell cycle oscillations
to growth arrest.}
(a) State transition graphs representing (left) the cyclic cellular dynamics corresponding
to cell division and (right) convergence to a globally attracting steady state ($100$) representing
a cell that has stopped dividing. The cellular states, shown as circles, are identified by
binary strings whose entries indicate if the molecular concentrations of the oscillator
components, viz., $A$, $B$
and $C$, respectively, are above a threshold value $K$. The cell switches from
the dynamics represented by one graph to that of the other when
the bifurcation parameter $k_8$, the strength of coupling between the contact-induced signal
and the cell cycle oscillator, is increased above the critical value  $k_{8}^{*}$.
(b) Comparison of the oscillations exhibited by the cell cycle model (top)
and those in the reduced model (bottom) obtained by replacing the continuous functions
describing the interactions between A, B and C with step functions.
(c) The scaled oscillation frequency $\nu^\prime$ (expressed relative to its maximum
possible value) shown as a function of the magnitude of the signal $S$ and the strength of
coupling $k_8$. The parameter space diagram for the cell cycle model (left) is seen
to be qualitatively similar to that obtained for the reduced model (center), which
in turn can be reproduced with high degree of accuracy using a closed-form
expression obtained analytically (right).
}
\label{figure2}
\end{figure}

The cell cycle model can be simplified further to make it analytically tractable by replacing
each of the continuously varying Hill functions with a Heaviside step function [$\Theta(z) = 1$ if $z \geq 0$, $=0$ otherwise] that exhibits
a discontinuous transition at a  threshold value of the argument.
The dynamics of the resulting \textit{reduced model} is
described by the following system of equations:
\begin{eqnarray*}
  \frac{dA}{dt} &=& [k_1 + k_7 \Theta(A-K)](A_T - A) - k_2 A \, \Theta(C-K), \\
  \frac{dB}{dt} &=& k_3 (B_T - B) - k_4 B \, \Theta(A-K) \,, \\
  \frac{dC}{dt} &=& k_5 (C_T - C) - k_6 C \, \Theta(B-K) - k_8 \frac{S^g}{\Psi^g + S^g} C.
\end{eqnarray*}
The dynamics of this system can be represented as trajectories between a set of discrete states
which are defined based upon whether the concentrations of A, B and C exceed the
threshold value $K$.
Thus we can represent each of the states in terms of binary strings of length $3$,
with the bit representing a particular molecular species being $0$ or $1$
if the corresponding concentration is  $<K$ or $>K$, respectively
[Fig.~\ref{figure2}~(a)].
Two different attractors are observed depending on the strength $k_8$ of the coupling
of the cell cycle oscillator to the signal S. We note that despite the differences in the pattern of oscillations exhibited by the cell cycle
model (Eqs.~\ref{eq:A}-\ref{eq:C}) and its reduced version [compare the two panels of
Fig.~\ref{figure2}~(b)], the curve $k_8^*(S)$
separating
the domain of these two attractors in the ($k_8,S$) parameter space are qualitatively similar
[compare the left and central panels of Fig.~\ref{figure2}~(c)].
For the reduced model, we can derive an exact expression for $k_8^* = k_5 [(C_T/K) - 1]/[S^g/(\psi^g + S^g])$, such that
for $k_8 < k_8^*$, 
the system periodically cycles between $6$ states, while for $k_8 >k_8^*$, the
dynamics converges to a fixed point when $S$ is sufficiently large.
In the former case, the oscillation period is the sum of the time intervals between the transitions
of the system from one state to another which comprise the cyclic attractor.
To obtain these intervals, we note that each of the transitions corresponds to any one of A, B and C
crossing the threshold $K$ either from below ($0 \rightarrow 1$) or above ($1 \rightarrow 0$).
Each transition $j$ is described by linear equations of the form
$\frac{dx_j}{dt} = \alpha_j - \beta_j x$, where $x_j$ represents the molecular concentration that
crosses the threshold in the transition $j$ and the parameters $\alpha$ and
$\beta$ are functions of the rate constants $k_1, \ldots k_8$, the total concentrations
$A_T, B_T, C_T$, the Hill function parameters $g$, $\psi$ and the strength of the
contact-induced signal $S$~\cite{SI}. 
Solving the equations, we obtain
the period as the sum of the time intervals $\tau$ required to switch from one state to another,
with those corresponding
to crossing the threshold from above being given by $\tau (0 \rightarrow 1) = -(1/\beta_j)
\log(1- \{\beta_j K/\alpha_j\})$ and those for crossing the threshold from below being
$\tau (1 \rightarrow 0) = -(1/\beta_j) \log(1- \{[\alpha_j/\beta_j] - K\}/\{[\alpha_j/\beta_j] - T\})$.
The frequency of oscillations obtained from this expression reproduces accurately the
results obtained by simulating the reduced model [compare the central and right panels of
Fig.~\ref{figure2}~(c)].

We note that the period of the cell cycle increases with the magnitude $S$ of the
contact-induced signal, and for large values of the coupling $k_8$, results in arrest
of the oscillations thereby halting further cell-division. 
To reproduce a CIP-like scenario with the increasing
cell density
expected after successive rounds of cell divisions,
we note that the cells become more likely to come in contact with each other over time,
thereby increasing $S$ on average.
This suggests that there is
an effective `negative feedback' operating between the rate at which cells multiply
governed by the cellular oscillator comprising the molecules A, B and C, and the resultant local
cell density which regulates the dynamics through the mechanism of contact-mediated
signaling. Such a process will result in the sequence shown in Fig.~\ref{figure1}~(b),
with cells initially proliferating rapidly, then slowing down over time and eventually cease
dividing altogether as their density progressively rises.
As cellular proliferation also serves to homogenize the cellular morphology in a
tightly packed domain, as is the case in growing epithelial tissue~\cite{Gibson2006,Lecuit2007,Devany2020}, a more intriguing possibility is that
changes in the sizes and shapes of cells comprising a confluent tissue
can themselves alter the contact-induced signal. As we show,
these morphological transitions can indeed control the periods of the cell cycles,
thereby influencing the rate at which cells proliferate.

\begin{figure}[htbp!]
\includegraphics{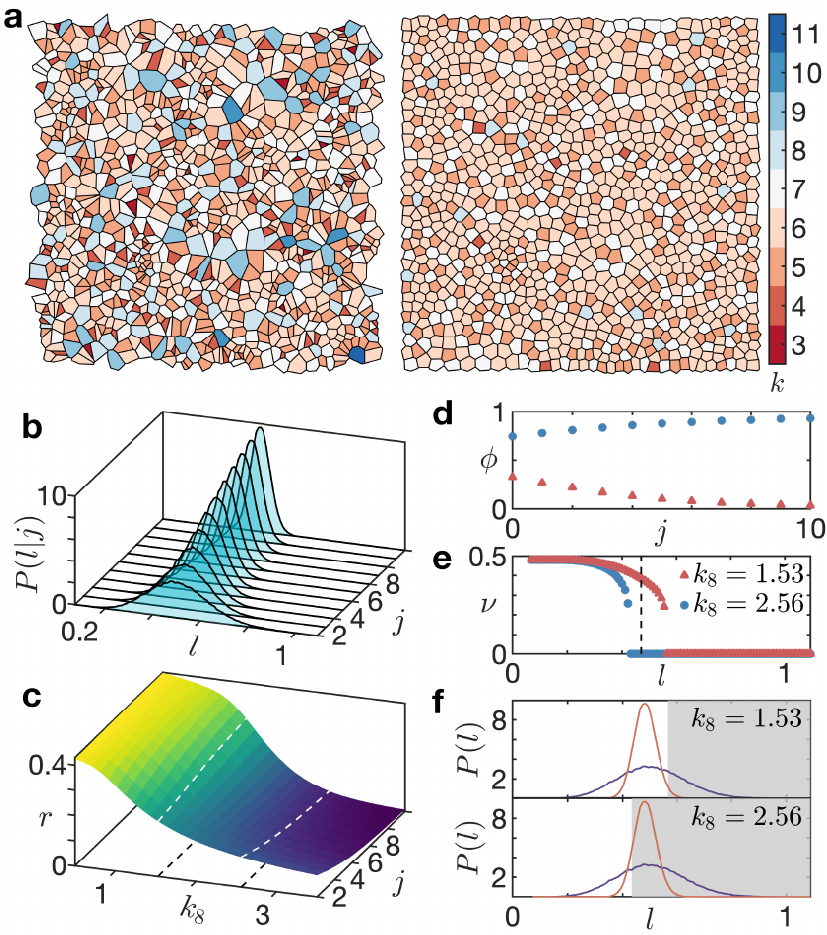}
\caption{\textbf{
Increasing homogeneity in the distribution of cell shapes can have differential
outcomes depending on the coupling between the contact-induced signal and the cell cycle
oscillator.}
(a) Representation of cellular packing in an epithelial sheet for different
levels of heterogeneity in the degree $k$, i.e., the number of cells
a given cell is in contact with
(indicated by the color of each polygon, see colorbar),
which is correlated to its area as well as perimeter, obtained by
Voronoi tessellation.
The initial, highly heterogeneous, configuration of $10^3$ cellular polygons (left panel) is
progressively homogenized by iterative application of Lloyd's algorithm, yielding
after $10$ iterations a tiling (right panel) that approximates
an optimal centroidal Voronoi tessellation (CVT), corresponding to
uniform cell density across the sheet. The square domain representing the tissue
has linear dimension of $4$ arbitrary length units.
(b) The approach to uniformity with successive iterations $j$ is indicated by
the evolution of the distribution of $l$, the perimeters of the cellular polygons tiling the sheet,
which is seen to become progressively narrower with $j$.
(c) The rate at which the tissue grows by cell division, $r$, given by the mean
of the frequencies of the cellular oscillators, varies with the heterogeneity in
a configuration (determined by the iteration $j$) and the bifurcation parameter $k_8$.
Two contrasting regimes are observed as the polygons become more uniform:
for lower values of $k_8$ the growth rate is seen to increase, while at higher
values, it decreases as the tissue becomes more homogeneous.
(d) The fraction $\phi$ of cells that have stopped oscillating at each iteration $j$
and (e) the frequency of oscillations of cells having perimeter $l$, shown for the
two regimes,  viz., $k_8 = 1.53$ (triangles) and $k_8 = 2.56$ (circles)
[corresponding to the dashed curves in (c)]. The broken line in (e) indicates the mean
perimeter of the $N$ cells tiling the tissue.
(f) The initial ($j=0$, blue) and final ($j=10$, red) distributions of $l$  corresponding
to the weak (lower panel: $k_8 = 1.53$) and strong coupling (upper panel: $k_8 = 2.56$)
regimes. The shaded region indicates perimeters above the critical value beyond which the oscillations in cells are arrested.
%
}
\label{figure3}
\end{figure}

To investigate how the rate of proliferation (controlled by the period of the cell cycle)
alters with the shape of the cells in a growing tissue, we consider a two-dimensional
plane tiled with cells whose shapes are approximated as polygonal.
Following Ref.~\cite{Sanchez2016}, we use Voronoi diagrams of
non-overlapping polygons to represent the space-filling arrangement of cells
in a tissue. Each polygon is associated with a
corresponding generating point or``seed'', such that its edges enclose all points to which this seed is the closest.
The polygons are initially obtained by randomly choosing $N$ generating points uniformly distributed across
the planar domain, which results in a highly heterogeneous distribution in terms of their number
of sides, areas and perimeters.
Subsequently we employ a stochastic version of Lloyd's algorithm to iteratively
generate progressively more homogeneous arrangements of these $N$ 
polygons~\cite{Sanchez2016}, reproducing the evolution of the distribution of
cellular geometries observed in normal development as cells proliferate (e.g., see
Refs.~\cite{Kokic2019,Dye2021})
without explicitly incorporating cell division in our model.
The process involves replacing the generating point of every Voronoi cell by an approximation
of their centroid (obtained as the mean of the coordinates of many randomly generated points inside
the polygon) at each iteration and recomputing the Voronoi diagram~\cite{note1}.
Applying this
sequence of steps repeatedly would eventually make the cellular arrangement converge to a
centroidal Voronoi tesselation (CVT) in which
the centroids and generating points coincide for all cells, corresponding to the most uniform
tiling of the domain using $N$ cells.
%
Fig.~\ref{figure3}~(a) shows a representative initial arrangement of cells that are highly
heterogeneous in terms of sizes and shapes (left panel). For comparison, we show alongside it
the configuration (right panel) obtained after $10$ iterations of the algorithm described above,
whose relatively higher homogeneity is visually apparent.
To demonstrate this quantitatively, Fig.~\ref{figure3}~(b) shows the evolution of the distribution
of the perimeters $l$ of the $N$ polygons tiling the plane through each step $j$ in the
the transition from the initial to the final state shown in Fig.~\ref{figure3}~(a).

The homogenization of cell sizes and shapes can in turn affect the rate at which the
tissue grows, as the frequency of ligand-receptor binding events that trigger the
signaling cascade regulating the cell cycle is  dependent on the perimeter of the cell.
Thus, we assume that the ligand $L_i$ bound to cell $i$ is proportional to its perimeter $l_i$,
viz., $L_i = (l_i/\langle l \rangle)L_0$, where $\langle l \rangle$ and $L_0$ are the average
perimeter of the cells and the mean ligand concentration across the tissue, respectively.
The corresponding strength $S_i$ of the contact-induced signal for the cell is given by
$S_i = S_{\max}~ L_i^q/(K_S^q + L_i^q)$, with
$S_{\max}$, $K_S$ and $q$ representing the maximum signal strength, the half-saturation constant
and the Hill coefficient regulating the steepness of the response function, respectively. We have explicitly verified that other possible dependences of the
signal on cell size/shape  (e.g., area or number of neighbors) yield results that
are qualitatively similar to those reported below~\cite{SI}.
Fig.~\ref{figure3}~(c) shows that the growth rate $r$ of the tissue (obtained by averaging
over the cell cycle frequencies across the domain) varies systematically as the cellular configuration
becomes more homogeneous (with increasing number of iterations $j$ of the Lloyd's algorithm).
However, there are two distinct regimes that are distinguished by the nature of this change
according to the value of
$k_8$, the strength of interaction between the contact-induced signal and the cell cycle oscillator.
For stronger $k_8$ [$ = 2.56$ in Fig.~\ref{figure3}~(c)]
increased homogeneity in cell shape and size is accompanied by
a slowing growth rate, while for weaker $k_8$ [$ = 1.53$ in Fig.~\ref{figure3}~(c)],
growth rate decreases with increasing heterogeneity.

To understand the genesis of the behavior seen in the two contrasting regimes, we
first note that with increasing homogeneity an increased fraction $\phi$
of cells have their oscillation arrested when the coupling $k_8$ is strong, while the reverse is
true for weaker $k_8$ [Fig.~\ref{figure3}~(d)]. This can be understood
in terms of the role that the cell perimeter $l$, which has a monotonic relation
to the magnitude of the contact-induced signal $S$, plays in the cell cycle oscillator.
Fig.~\ref{figure3}~(e) shows that while increased $l$ results in all cases in the
oscillation frequency decreasing eventually to $0$, the critical cell size at which
the oscillation is arrested is lowered as the coupling $k_8$ becomes stronger
[consistent with Fig.~\ref{figure2}~(c)]. As increased homogeneity implies a decreasing width of the distribution of $l$, we can now
explain the differential evolution of the tissue growth rate in the two regimes as follows.
For weaker $k_8$, where the critical
value of $l$  (indicated by the boundary of the shaded region) is higher than the mean perimeter, with decreasing width of the distribution we will
observe a relative {\em increase} in the fraction of oscillating cells and also their mean frequency [Fig.~\ref{figure3}~(f), upper panel]. It follows that decreasing growth rate will be associated with
increasing {\em heterogeneity} in cell shapes and sizes in the weak coupling regime.
In contrast, the lower panel in Fig.~\ref{figure3}~(f) shows that for a stronger value of $k_8$, when the
critical value of $l$ at which oscillation is arrested is lower than the mean cellular
perimeter, the fraction of oscillating cells (as well as their mean frequency) will {\em decrease}
as the dispersion of $l$ decreases. Thus, in this
strong coupling regime, it is increased {\em homogeneity} of cellular morphology that will accompany slowing growth rate of the tissue.

To conclude, we have shown that increasing heterogeneity in cell sizes and shapes can lead to differential outcomes in the collective activity of a system of cell cycle oscillators arranged on a disordered lattice, depending on the strength of the inter-cellular interactions that implement contact inhibition.
With decreasing role of the contact inhibition signal in modulating the cell cycle oscillator,
i.e., at low $k_8$, we expect the growth rate to increase as the  cellular arrangement
becomes more irregular, as is observed in dysplasia that sometimes precedes tumor growth.
Indeed, inter-cellular communication is known to be impeded in cancers
along with increased rate of cellular proliferation~\cite{Wada2011,Knights2012,Yu2016},
which can be associated in our model to a system trajectory involving both $k_8$ (regulating the signaling between cells) and $j$ (controlling
the disorder in cellular morphology) decreasing simultaneously,
causing the proliferation rate $r$ to increase.
In the high $k_8$ regime where the dynamics of the cell cycle oscillator is more sensitive to
contact-induced signals, increased homogenization in the contact topology
of cells leads to reduction of the fraction of oscillating cells, as well as their
mean frequency, which together contribute to the growth rate.
Thus, our results suggest a causal relation between
the observation of the simultaneous increase in regularity
in the planar arrangement of cells in growing epithelial tissue and the
arrest of growth in the cellular assembly, that occur together
over the normal course of development~\cite{Gibson2006,Classen2005,note3}.
Indeed, it hints that, in general, heterogeneous contact topology
in networks of oscillators interacting via lateral inhibition~\cite{Janaki2019}
will increase the range of interaction strengths over which chimera
states, characterized by coexistence of oscillating and non-oscillating
elements~\cite{Singh2013}, are likely to appear.
\begin{acknowledgments}
The authors would like to acknowledge discussions during the ICTS Winter
School on Quantitative Systems Biology (ICTS/qsb2019/12).
SNM has been supported by the Center of Excellence in Complex Systems and Data
Science, funded by the Department of Atomic Energy, Government of
India. The simulations required for this work were
supported by IMSc High Performance Computing facility (hpc.imsc.res.in) [Nandadevi].
\end{acknowledgments}


\begin{thebibliography}{99}
\makeatletter
\providecommand \@ifxundefined [1]{%
 \@ifx{#1\undefined}
}%
\providecommand \@ifnum [1]{%
 \ifnum #1\expandafter \@firstoftwo
 \else \expandafter \@secondoftwo
 \fi
}%
\providecommand \@ifx [1]{%
 \ifx #1\expandafter \@firstoftwo
 \else \expandafter \@secondoftwo
 \fi
}%
\providecommand \natexlab [1]{#1}%
\providecommand \enquote  [1]{``#1''}%
\providecommand \bibnamefont  [1]{#1}%
\providecommand \bibfnamefont [1]{#1}%
\providecommand \citenamefont [1]{#1}%
\providecommand \href@noop [0]{\@secondoftwo}%
\providecommand \href [0]{\begingroup \@sanitize@url \@href}%
\providecommand \@href[1]{\@@startlink{#1}\@@href}%
\providecommand \@@href[1]{\endgroup#1\@@endlink}%
\providecommand \@sanitize@url [0]{\catcode `\\12\catcode `\$12\catcode
  `\&12\catcode `\#12\catcode `\^12\catcode `\_12\catcode `\%12\relax}%
\providecommand \@@startlink[1]{}%
\providecommand \@@endlink[0]{}%
\providecommand \url  [0]{\begingroup\@sanitize@url \@url }%
\providecommand \@url [1]{\endgroup\@href {#1}{\urlprefix }}%
\providecommand \urlprefix  [0]{URL }%
\providecommand \Eprint [0]{\href }%
\providecommand \doibase [0]{http://dx.doi.org/}%
\providecommand \selectlanguage [0]{\@gobble}%
\providecommand \bibinfo  [0]{\@secondoftwo}%
\providecommand \bibfield  [0]{\@secondoftwo}%
\providecommand \translation [1]{[#1]}%
\providecommand \BibitemOpen [0]{}%
\providecommand \bibitemStop [0]{}%
\providecommand \bibitemNoStop [0]{.\EOS\space}%
\providecommand \EOS [0]{\spacefactor3000\relax}%
\providecommand \BibitemShut  [1]{\csname bibitem#1\endcsname}%
\let\auto@bib@innerbib\@empty
\bibitem [{\citenamefont {Gilbert}(2018)}]{Gilbert2018}%
  \BibitemOpen
  \bibfield  {author} {\bibinfo {author} {\bibfnamefont {S.}~\bibnamefont
  {Gilbert}},\ }\href@noop {} {\emph {\bibinfo {title} {{Developmental Biology,
  11th ed.}}}}\ (\bibinfo  {publisher} {Sinauer},\ \bibinfo {address}
  {Sunderland, MA},\ \bibinfo {year} {2018})\BibitemShut {NoStop}%
\bibitem [{\citenamefont {Wolpert}\ and\ \citenamefont
  {Tickle}(2011)}]{Wolpert2011}%
  \BibitemOpen
  \bibfield  {author} {\bibinfo {author} {\bibfnamefont {L.}~\bibnamefont
  {Wolpert}}\ and\ \bibinfo {author} {\bibfnamefont {C.}~\bibnamefont
  {Tickle}},\ }\href@noop {} {\emph {\bibinfo {title} {{Developmental
  Biology}}}}\ (\bibinfo  {publisher} {Oxford Univ. Press},\ \bibinfo {address}
  {Oxford},\ \bibinfo {year} {2011})\BibitemShut {NoStop}%
\bibitem [{\citenamefont {Shahbazi}(2020)}]{Shahbazi2020}%
  \BibitemOpen
  \bibfield  {author} {\bibinfo {author} {\bibfnamefont {M.~N.}\ \bibnamefont
  {Shahbazi}},\ }\href {\doibase 10.1242/dev.190629} {\bibfield  {journal}
  {\bibinfo  {journal} {Development}\ }\textbf {\bibinfo {volume} {147}}
  (\bibinfo {year} {2020}),\ 10.1242/dev.190629}\BibitemShut {NoStop}%
\bibitem [{\citenamefont {De~Monte}\ \emph {et~al.}(2007)\citenamefont
  {De~Monte}, \citenamefont {d'Ovidio}, \citenamefont {Dan{\o}},\ and\
  \citenamefont {S{\o}rensen}}]{DeMonte2007}%
  \BibitemOpen
  \bibfield  {author} {\bibinfo {author} {\bibfnamefont {S.}~\bibnamefont
  {De~Monte}}, \bibinfo {author} {\bibfnamefont {F.}~\bibnamefont {d'Ovidio}},
  \bibinfo {author} {\bibfnamefont {S.}~\bibnamefont {Dan{\o}}}, \ and\
  \bibinfo {author} {\bibfnamefont {P.~G.}\ \bibnamefont {S{\o}rensen}},\
  }\href {\doibase 10.1073/pnas.0706089104} {\bibfield  {journal} {\bibinfo
  {journal} {Proc. Natl. Acad. Sci. USA}\ }\textbf {\bibinfo {volume} {104}},\
  \bibinfo {pages} {18377} (\bibinfo {year} {2007})}\BibitemShut {NoStop}%
\bibitem [{\citenamefont {Taylor}\ \emph {et~al.}(2009)\citenamefont {Taylor},
  \citenamefont {Tinsley}, \citenamefont {Wang}, \citenamefont {Huang},\ and\
  \citenamefont {Showalter}}]{Taylor2009}%
  \BibitemOpen
  \bibfield  {author} {\bibinfo {author} {\bibfnamefont {A.~F.}\ \bibnamefont
  {Taylor}}, \bibinfo {author} {\bibfnamefont {M.~R.}\ \bibnamefont {Tinsley}},
  \bibinfo {author} {\bibfnamefont {F.}~\bibnamefont {Wang}}, \bibinfo {author}
  {\bibfnamefont {Z.}~\bibnamefont {Huang}}, \ and\ \bibinfo {author}
  {\bibfnamefont {K.}~\bibnamefont {Showalter}},\ }\href {\doibase
  10.1126/science.1166253} {\bibfield  {journal} {\bibinfo  {journal}
  {Science}\ }\textbf {\bibinfo {volume} {323}},\ \bibinfo {pages} {614}
  (\bibinfo {year} {2009})}\BibitemShut {NoStop}%
\bibitem [{\citenamefont {Whiteley}\ \emph {et~al.}(2017)\citenamefont
  {Whiteley}, \citenamefont {Diggle},\ and\ \citenamefont
  {Greenberg}}]{Whiteley2017}%
  \BibitemOpen
  \bibfield  {author} {\bibinfo {author} {\bibfnamefont {M.}~\bibnamefont
  {Whiteley}}, \bibinfo {author} {\bibfnamefont {S.~P.}\ \bibnamefont
  {Diggle}}, \ and\ \bibinfo {author} {\bibfnamefont {E.~P.}\ \bibnamefont
  {Greenberg}},\ }\href {\doibase 10.1038/nature24624} {\bibfield  {journal}
  {\bibinfo  {journal} {Nature (London)}\ }\textbf {\bibinfo {volume} {551}},\
  \bibinfo {pages} {313} (\bibinfo {year} {2017})}\BibitemShut {NoStop}%
\bibitem [{\citenamefont {Lander}(2013)}]{Lander2013}%
  \BibitemOpen
  \bibfield  {author} {\bibinfo {author} {\bibfnamefont {A.~D.}\ \bibnamefont
  {Lander}},\ }\href {\doibase 10.1126/science.1224186} {\bibfield  {journal}
  {\bibinfo  {journal} {Science}\ }\textbf {\bibinfo {volume} {339}},\ \bibinfo
  {pages} {923} (\bibinfo {year} {2013})}\BibitemShut {NoStop}%
\bibitem [{\citenamefont {Kuyyamudi}\ \emph {et~al.}(2021)\citenamefont
  {Kuyyamudi}, \citenamefont {Menon},\ and\ \citenamefont
  {Sinha}}]{Kuyyamudi2021}%
  \BibitemOpen
  \bibfield  {author} {\bibinfo {author} {\bibfnamefont {C.}~\bibnamefont
  {Kuyyamudi}}, \bibinfo {author} {\bibfnamefont {S.~N.}\ \bibnamefont
  {Menon}}, \ and\ \bibinfo {author} {\bibfnamefont {S.}~\bibnamefont
  {Sinha}},\ }\href {\doibase 10.1103/PhysRevE.103.062409} {\bibfield
  {journal} {\bibinfo  {journal} {Phys. Rev. E}\ }\textbf {\bibinfo {volume}
  {103}},\ \bibinfo {pages} {062409} (\bibinfo {year} {2021})}\BibitemShut
  {NoStop}%
\bibitem [{\citenamefont {Hagan}\ and\ \citenamefont
  {Grason}(2021)}]{Hagan2021}%
  \BibitemOpen
  \bibfield  {author} {\bibinfo {author} {\bibfnamefont {M.~F.}\ \bibnamefont
  {Hagan}}\ and\ \bibinfo {author} {\bibfnamefont {G.~M.}\ \bibnamefont
  {Grason}},\ }\href {\doibase 10.1103/RevModPhys.93.025008} {\bibfield
  {journal} {\bibinfo  {journal} {Rev. Mod. Phys.}\ }\textbf {\bibinfo {volume}
  {93}},\ \bibinfo {pages} {025008} (\bibinfo {year} {2021})}\BibitemShut
  {NoStop}%
\bibitem [{\citenamefont {Sherr}(1996)}]{Sherr1996}%
  \BibitemOpen
  \bibfield  {author} {\bibinfo {author} {\bibfnamefont {C.~J.}\ \bibnamefont
  {Sherr}},\ }\href {\doibase 10.1126/science.274.5293.1672} {\bibfield
  {journal} {\bibinfo  {journal} {Science}\ }\textbf {\bibinfo {volume}
  {274}},\ \bibinfo {pages} {1672} (\bibinfo {year} {1996})}\BibitemShut
  {NoStop}%
\bibitem [{\citenamefont {Kuznetsov}\ and\ \citenamefont
  {Blokhin}(2000)}]{Kuznetsov2000}%
  \BibitemOpen
  \bibfield  {author} {\bibinfo {author} {\bibfnamefont {D.~V.}\ \bibnamefont
  {Kuznetsov}}\ and\ \bibinfo {author} {\bibfnamefont {A.~V.}\ \bibnamefont
  {Blokhin}},\ }\href {\doibase 10.1103/PhysRevLett.85.2833} {\bibfield
  {journal} {\bibinfo  {journal} {Phys. Rev. Lett.}\ }\textbf {\bibinfo
  {volume} {85}},\ \bibinfo {pages} {2833} (\bibinfo {year}
  {2000})}\BibitemShut {NoStop}%
\bibitem [{\citenamefont {Santamar{\'\i}a}\ \emph {et~al.}(2007)\citenamefont
  {Santamar{\'\i}a}, \citenamefont {Barri{\`e}re}, \citenamefont {Cerqueira},
  \citenamefont {Hunt}, \citenamefont {Tardy}, \citenamefont {Newton},
  \citenamefont {C{\'a}ceres}, \citenamefont {Dubus}, \citenamefont
  {Malumbres},\ and\ \citenamefont {Barbacid}}]{Santamaria2007}%
  \BibitemOpen
  \bibfield  {author} {\bibinfo {author} {\bibfnamefont {D.}~\bibnamefont
  {Santamar{\'\i}a}}, \bibinfo {author} {\bibfnamefont {C.}~\bibnamefont
  {Barri{\`e}re}}, \bibinfo {author} {\bibfnamefont {A.}~\bibnamefont
  {Cerqueira}}, \bibinfo {author} {\bibfnamefont {S.}~\bibnamefont {Hunt}},
  \bibinfo {author} {\bibfnamefont {C.}~\bibnamefont {Tardy}}, \bibinfo
  {author} {\bibfnamefont {K.}~\bibnamefont {Newton}}, \bibinfo {author}
  {\bibfnamefont {J.~F.}\ \bibnamefont {C{\'a}ceres}}, \bibinfo {author}
  {\bibfnamefont {P.}~\bibnamefont {Dubus}}, \bibinfo {author} {\bibfnamefont
  {M.}~\bibnamefont {Malumbres}}, \ and\ \bibinfo {author} {\bibfnamefont
  {M.}~\bibnamefont {Barbacid}},\ }\href {\doibase 10.1038/nature06046}
  {\bibfield  {journal} {\bibinfo  {journal} {Nature (London)}\ }\textbf
  {\bibinfo {volume} {448}},\ \bibinfo {pages} {811} (\bibinfo {year}
  {2007})}\BibitemShut {NoStop}%
\bibitem [{\citenamefont {Orlando}\ \emph {et~al.}(2008)\citenamefont
  {Orlando}, \citenamefont {Lin}, \citenamefont {Bernard}, \citenamefont
  {Wang}, \citenamefont {Socolar}, \citenamefont {Iversen}, \citenamefont
  {Hartemink},\ and\ \citenamefont {Haase}}]{Orlando2008}%
  \BibitemOpen
  \bibfield  {author} {\bibinfo {author} {\bibfnamefont {D.~A.}\ \bibnamefont
  {Orlando}}, \bibinfo {author} {\bibfnamefont {C.~Y.}\ \bibnamefont {Lin}},
  \bibinfo {author} {\bibfnamefont {A.}~\bibnamefont {Bernard}}, \bibinfo
  {author} {\bibfnamefont {J.~Y.}\ \bibnamefont {Wang}}, \bibinfo {author}
  {\bibfnamefont {J.~E.}\ \bibnamefont {Socolar}}, \bibinfo {author}
  {\bibfnamefont {E.~S.}\ \bibnamefont {Iversen}}, \bibinfo {author}
  {\bibfnamefont {A.~J.}\ \bibnamefont {Hartemink}}, \ and\ \bibinfo {author}
  {\bibfnamefont {S.~B.}\ \bibnamefont {Haase}},\ }\href {\doibase
  10.1038/nature06955} {\bibfield  {journal} {\bibinfo  {journal} {Nature
  (London)}\ }\textbf {\bibinfo {volume} {453}},\ \bibinfo {pages} {944}
  (\bibinfo {year} {2008})}\BibitemShut {NoStop}%
\bibitem [{\citenamefont {Stoker}\ and\ \citenamefont
  {Rubin}(1967)}]{Stoker1967}%
  \BibitemOpen
  \bibfield  {author} {\bibinfo {author} {\bibfnamefont {M.}~\bibnamefont
  {Stoker}}\ and\ \bibinfo {author} {\bibfnamefont {H.}~\bibnamefont {Rubin}},\
  }\href {\doibase 10.1038/215171a0} {\bibfield  {journal} {\bibinfo  {journal}
  {Nature (London)}\ }\textbf {\bibinfo {volume} {215}},\ \bibinfo {pages}
  {171} (\bibinfo {year} {1967})}\BibitemShut {NoStop}%
\bibitem [{\citenamefont {{\v{S}}estan}\ \emph {et~al.}(1999)\citenamefont
  {{\v{S}}estan}, \citenamefont {Artavanis-Tsakonas},\ and\ \citenamefont
  {Rakic}}]{Sestan1999}%
  \BibitemOpen
  \bibfield  {author} {\bibinfo {author} {\bibfnamefont {N.}~\bibnamefont
  {{\v{S}}estan}}, \bibinfo {author} {\bibfnamefont {S.}~\bibnamefont
  {Artavanis-Tsakonas}}, \ and\ \bibinfo {author} {\bibfnamefont
  {P.}~\bibnamefont {Rakic}},\ }\href {\doibase 10.1126/science.286.5440.741}
  {\bibfield  {journal} {\bibinfo  {journal} {Science}\ }\textbf {\bibinfo
  {volume} {286}},\ \bibinfo {pages} {741} (\bibinfo {year}
  {1999})}\BibitemShut {NoStop}%
\bibitem [{\citenamefont {Zhao}\ \emph {et~al.}(2007)\citenamefont {Zhao},
  \citenamefont {Wei}, \citenamefont {Li}, \citenamefont {Udan}, \citenamefont
  {Yang}, \citenamefont {Kim}, \citenamefont {Xie}, \citenamefont {Ikenoue},
  \citenamefont {Yu}, \citenamefont {Li} \emph {et~al.}}]{Zhao2007}%
  \BibitemOpen
  \bibfield  {author} {\bibinfo {author} {\bibfnamefont {B.}~\bibnamefont
  {Zhao}}, \bibinfo {author} {\bibfnamefont {X.}~\bibnamefont {Wei}}, \bibinfo
  {author} {\bibfnamefont {W.}~\bibnamefont {Li}}, \bibinfo {author}
  {\bibfnamefont {R.~S.}\ \bibnamefont {Udan}}, \bibinfo {author}
  {\bibfnamefont {Q.}~\bibnamefont {Yang}}, \bibinfo {author} {\bibfnamefont
  {J.}~\bibnamefont {Kim}}, \bibinfo {author} {\bibfnamefont {J.}~\bibnamefont
  {Xie}}, \bibinfo {author} {\bibfnamefont {T.}~\bibnamefont {Ikenoue}},
  \bibinfo {author} {\bibfnamefont {J.}~\bibnamefont {Yu}}, \bibinfo {author}
  {\bibfnamefont {L.}~\bibnamefont {Li}},  \emph {et~al.},\ }\href {\doibase
  10.1101/gad.1602907} {\bibfield  {journal} {\bibinfo  {journal} {Genes Dev.}\
  }\textbf {\bibinfo {volume} {21}},\ \bibinfo {pages} {2747} (\bibinfo {year}
  {2007})}\BibitemShut {NoStop}%
\bibitem [{\citenamefont {Kim}\ \emph {et~al.}(2011)\citenamefont {Kim},
  \citenamefont {Koh}, \citenamefont {Chen},\ and\ \citenamefont
  {Gumbiner}}]{Kim2011}%
  \BibitemOpen
  \bibfield  {author} {\bibinfo {author} {\bibfnamefont {N.-G.}\ \bibnamefont
  {Kim}}, \bibinfo {author} {\bibfnamefont {E.}~\bibnamefont {Koh}}, \bibinfo
  {author} {\bibfnamefont {X.}~\bibnamefont {Chen}}, \ and\ \bibinfo {author}
  {\bibfnamefont {B.~M.}\ \bibnamefont {Gumbiner}},\ }\href {\doibase
  10.1073/pnas.1103345108} {\bibfield  {journal} {\bibinfo  {journal} {Proc.
  Natl. Acad. Sci. USA}\ }\textbf {\bibinfo {volume} {108}},\ \bibinfo {pages}
  {11930} (\bibinfo {year} {2011})}\BibitemShut {NoStop}%
\bibitem [{\citenamefont {Puliafito}\ \emph {et~al.}(2012)\citenamefont
  {Puliafito}, \citenamefont {Hufnagel}, \citenamefont {Neveu}, \citenamefont
  {Streichan}, \citenamefont {Sigal}, \citenamefont {Fygenson},\ and\
  \citenamefont {Shraiman}}]{Puliafito2012}%
  \BibitemOpen
  \bibfield  {author} {\bibinfo {author} {\bibfnamefont {A.}~\bibnamefont
  {Puliafito}}, \bibinfo {author} {\bibfnamefont {L.}~\bibnamefont {Hufnagel}},
  \bibinfo {author} {\bibfnamefont {P.}~\bibnamefont {Neveu}}, \bibinfo
  {author} {\bibfnamefont {S.}~\bibnamefont {Streichan}}, \bibinfo {author}
  {\bibfnamefont {A.}~\bibnamefont {Sigal}}, \bibinfo {author} {\bibfnamefont
  {D.~K.}\ \bibnamefont {Fygenson}}, \ and\ \bibinfo {author} {\bibfnamefont
  {B.~I.}\ \bibnamefont {Shraiman}},\ }\href {\doibase 10.1073/pnas.1007809109}
  {\bibfield  {journal} {\bibinfo  {journal} {Proc. Natl. Acad. Sci. USA}\
  }\textbf {\bibinfo {volume} {109}},\ \bibinfo {pages} {739} (\bibinfo {year}
  {2012})}\BibitemShut {NoStop}%
\bibitem [{\citenamefont {Aoki}\ \emph {et~al.}(2013)\citenamefont {Aoki},
  \citenamefont {Kumagai}, \citenamefont {Sakurai}, \citenamefont {Komatsu},
  \citenamefont {Fujita}, \citenamefont {Shionyu},\ and\ \citenamefont
  {Matsuda}}]{Aoki2013}%
  \BibitemOpen
  \bibfield  {author} {\bibinfo {author} {\bibfnamefont {K.}~\bibnamefont
  {Aoki}}, \bibinfo {author} {\bibfnamefont {Y.}~\bibnamefont {Kumagai}},
  \bibinfo {author} {\bibfnamefont {A.}~\bibnamefont {Sakurai}}, \bibinfo
  {author} {\bibfnamefont {N.}~\bibnamefont {Komatsu}}, \bibinfo {author}
  {\bibfnamefont {Y.}~\bibnamefont {Fujita}}, \bibinfo {author} {\bibfnamefont
  {C.}~\bibnamefont {Shionyu}}, \ and\ \bibinfo {author} {\bibfnamefont
  {M.}~\bibnamefont {Matsuda}},\ }\href {\doibase 10.1016/j.molcel.2013.09.015}
  {\bibfield  {journal} {\bibinfo  {journal} {Mol. Cell}\ }\textbf {\bibinfo
  {volume} {52}},\ \bibinfo {pages} {529} (\bibinfo {year} {2013})}\BibitemShut
  {NoStop}%
\bibitem [{\citenamefont {Honda}(2017)}]{Honda2017}%
  \BibitemOpen
  \bibfield  {author} {\bibinfo {author} {\bibfnamefont {H.}~\bibnamefont
  {Honda}},\ }\href {\doibase 10.1111/dgd.12350} {\bibfield  {journal}
  {\bibinfo  {journal} {Dev. Growth Differ.}\ }\textbf {\bibinfo {volume}
  {59}},\ \bibinfo {pages} {306} (\bibinfo {year} {2017})}\BibitemShut
  {NoStop}%
\bibitem [{\citenamefont {McCaffrey}\ and\ \citenamefont
  {Macara}(2011)}]{Mccaffrey2011}%
  \BibitemOpen
  \bibfield  {author} {\bibinfo {author} {\bibfnamefont {L.~M.}\ \bibnamefont
  {McCaffrey}}\ and\ \bibinfo {author} {\bibfnamefont {I.~G.}\ \bibnamefont
  {Macara}},\ }\href {\doibase 10.1016/j.tcb.2011.06.005} {\bibfield  {journal}
  {\bibinfo  {journal} {Trends Cell Biol.}\ }\textbf {\bibinfo {volume} {21}},\
  \bibinfo {pages} {727} (\bibinfo {year} {2011})}\BibitemShut {NoStop}%
\bibitem [{\citenamefont {Macara}\ \emph {et~al.}(2014)\citenamefont {Macara},
  \citenamefont {Guyer}, \citenamefont {Richardson}, \citenamefont {Huo},\ and\
  \citenamefont {Ahmed}}]{Macara2014}%
  \BibitemOpen
  \bibfield  {author} {\bibinfo {author} {\bibfnamefont {I.~G.}\ \bibnamefont
  {Macara}}, \bibinfo {author} {\bibfnamefont {R.}~\bibnamefont {Guyer}},
  \bibinfo {author} {\bibfnamefont {G.}~\bibnamefont {Richardson}}, \bibinfo
  {author} {\bibfnamefont {Y.}~\bibnamefont {Huo}}, \ and\ \bibinfo {author}
  {\bibfnamefont {S.~M.}\ \bibnamefont {Ahmed}},\ }\href {\doibase
  10.1016/j.cub.2014.06.068} {\bibfield  {journal} {\bibinfo  {journal} {Curr.
  Biol.}\ }\textbf {\bibinfo {volume} {24}},\ \bibinfo {pages} {R815} (\bibinfo
  {year} {2014})}\BibitemShut {NoStop}%
\bibitem [{\citenamefont {Harvey}\ \emph {et~al.}(2003)\citenamefont {Harvey},
  \citenamefont {Pfleger},\ and\ \citenamefont {Hariharan}}]{Harvey2003}%
  \BibitemOpen
  \bibfield  {author} {\bibinfo {author} {\bibfnamefont {K.~F.}\ \bibnamefont
  {Harvey}}, \bibinfo {author} {\bibfnamefont {C.~M.}\ \bibnamefont {Pfleger}},
  \ and\ \bibinfo {author} {\bibfnamefont {I.~K.}\ \bibnamefont {Hariharan}},\
  }\href {\doibase 10.1016/s0092-8674(03)00557-9} {\bibfield  {journal}
  {\bibinfo  {journal} {Cell}\ }\textbf {\bibinfo {volume} {114}},\ \bibinfo
  {pages} {457} (\bibinfo {year} {2003})}\BibitemShut {NoStop}%
\bibitem [{\citenamefont {Huang}\ \emph {et~al.}(2005)\citenamefont {Huang},
  \citenamefont {Wu}, \citenamefont {Barrera}, \citenamefont {Matthews},\ and\
  \citenamefont {Pan}}]{Huang2005}%
  \BibitemOpen
  \bibfield  {author} {\bibinfo {author} {\bibfnamefont {J.}~\bibnamefont
  {Huang}}, \bibinfo {author} {\bibfnamefont {S.}~\bibnamefont {Wu}}, \bibinfo
  {author} {\bibfnamefont {J.}~\bibnamefont {Barrera}}, \bibinfo {author}
  {\bibfnamefont {K.}~\bibnamefont {Matthews}}, \ and\ \bibinfo {author}
  {\bibfnamefont {D.}~\bibnamefont {Pan}},\ }\href {\doibase
  10.1016/j.cell.2005.06.007} {\bibfield  {journal} {\bibinfo  {journal}
  {Cell}\ }\textbf {\bibinfo {volume} {122}},\ \bibinfo {pages} {421} (\bibinfo
  {year} {2005})}\BibitemShut {NoStop}%
\bibitem [{\citenamefont {Thompson}\ and\ \citenamefont
  {Cohen}(2006)}]{Thompson2006}%
  \BibitemOpen
  \bibfield  {author} {\bibinfo {author} {\bibfnamefont {B.~J.}\ \bibnamefont
  {Thompson}}\ and\ \bibinfo {author} {\bibfnamefont {S.~M.}\ \bibnamefont
  {Cohen}},\ }\href {\doibase 10.1016/j.cell.2006.07.013} {\bibfield  {journal}
  {\bibinfo  {journal} {Cell}\ }\textbf {\bibinfo {volume} {126}},\ \bibinfo
  {pages} {767} (\bibinfo {year} {2006})}\BibitemShut {NoStop}%
\bibitem [{\citenamefont {Zeng}\ and\ \citenamefont {Hong}(2008)}]{Zeng2008}%
  \BibitemOpen
  \bibfield  {author} {\bibinfo {author} {\bibfnamefont {Q.}~\bibnamefont
  {Zeng}}\ and\ \bibinfo {author} {\bibfnamefont {W.}~\bibnamefont {Hong}},\
  }\href {\doibase 10.1016/j.ccr.2008.02.011} {\bibfield  {journal} {\bibinfo
  {journal} {Cancer Cell}\ }\textbf {\bibinfo {volume} {13}},\ \bibinfo {pages}
  {188} (\bibinfo {year} {2008})}\BibitemShut {NoStop}%
\bibitem [{\citenamefont {Kango-Singh}\ and\ \citenamefont
  {Singh}(2009)}]{Kango2009}%
  \BibitemOpen
  \bibfield  {author} {\bibinfo {author} {\bibfnamefont {M.}~\bibnamefont
  {Kango-Singh}}\ and\ \bibinfo {author} {\bibfnamefont {A.}~\bibnamefont
  {Singh}},\ }\href {\doibase 10.1002/dvdy.21996} {\bibfield  {journal}
  {\bibinfo  {journal} {Dev. Dyn.}\ }\textbf {\bibinfo {volume} {238}},\
  \bibinfo {pages} {1627} (\bibinfo {year} {2009})}\BibitemShut {NoStop}%
\bibitem [{\citenamefont {Oh}\ and\ \citenamefont {Irvine}(2010)}]{Oh2010}%
  \BibitemOpen
  \bibfield  {author} {\bibinfo {author} {\bibfnamefont {H.}~\bibnamefont
  {Oh}}\ and\ \bibinfo {author} {\bibfnamefont {K.~D.}\ \bibnamefont
  {Irvine}},\ }\href {\doibase 10.1016/j.tcb.2010.04.005} {\bibfield  {journal}
  {\bibinfo  {journal} {Trends Cell Biol.}\ }\textbf {\bibinfo {volume} {20}},\
  \bibinfo {pages} {410} (\bibinfo {year} {2010})}\BibitemShut {NoStop}%
\bibitem [{\citenamefont {Gumbiner}\ and\ \citenamefont
  {Kim}(2014)}]{Gumbiner2014}%
  \BibitemOpen
  \bibfield  {author} {\bibinfo {author} {\bibfnamefont {B.~M.}\ \bibnamefont
  {Gumbiner}}\ and\ \bibinfo {author} {\bibfnamefont {N.-G.}\ \bibnamefont
  {Kim}},\ }\href {\doibase 10.1242/dev.109108} {\bibfield  {journal} {\bibinfo
   {journal} {J. Cell Sci.}\ }\textbf {\bibinfo {volume} {127}},\ \bibinfo
  {pages} {709} (\bibinfo {year} {2014})}\BibitemShut {NoStop}%
 \bibitem{Ma2019}
\BibitemOpen
  \bibfield  {author} {
  \bibinfo {author} {\bibfnamefont {S.}\ \bibnamefont {Ma}},\
  \bibinfo {author} {\bibfnamefont {Z.}\ \bibnamefont {Meng}},\
  \bibinfo {author} {\bibfnamefont {R.}\ \bibnamefont {Chen}},\ and\
  \bibinfo {author} {\bibfnamefont {K.-L.}\ \bibnamefont {Guan}}, \ }
  \href {\doibase 10.1146/annurev-biochem-013118-111829}
  {\bibfield  {journal} {\bibinfo {journal} {Annu. Rev. Biochem.}\ }
   \textbf {\bibinfo {volume} {88}},\
   \bibinfo {pages} {577}
   (\bibinfo {year} {2019})}
   \BibitemShut {NoStop}%
\bibitem [{\citenamefont {Tsai}\ \emph {et~al.}(2008)\citenamefont {Tsai},
  \citenamefont {Choi}, \citenamefont {Ma}, \citenamefont {Pomerening},
  \citenamefont {Tang},\ and\ \citenamefont {Ferrell}}]{Tsai2008}%
  \BibitemOpen
  \bibfield  {author} {\bibinfo {author} {\bibfnamefont {T.~Y.-C.}\
  \bibnamefont {Tsai}}, \bibinfo {author} {\bibfnamefont {Y.~S.}\ \bibnamefont
  {Choi}}, \bibinfo {author} {\bibfnamefont {W.}~\bibnamefont {Ma}}, \bibinfo
  {author} {\bibfnamefont {J.~R.}\ \bibnamefont {Pomerening}}, \bibinfo
  {author} {\bibfnamefont {C.}~\bibnamefont {Tang}}, \ and\ \bibinfo {author}
  {\bibfnamefont {J.~E.}\ \bibnamefont {Ferrell}},\ }\href {\doibase
  10.1126/science.1156951} {\bibfield  {journal} {\bibinfo  {journal}
  {Science}\ }\textbf {\bibinfo {volume} {321}},\ \bibinfo {pages} {126}
  (\bibinfo {year} {2008})}\BibitemShut {NoStop}%
\bibitem{note2} 
Qualitatively similar behavior is obtained if $S$ is coupled to the cell-cycle
oscillator via $A$ or $B$ (instead of $C$) with the only change occurring in the
identity of the state to which the system converges ($010$ and $001$, respectively,
instead of $100$) for $k_8 \geq k_8^*$.%
\bibitem{SI}
 See Supplemental Material for details.
\bibitem [{\citenamefont {Gibson}\ \emph {et~al.}(2006)\citenamefont {Gibson},
  \citenamefont {Patel}, \citenamefont {Nagpal},\ and\ \citenamefont
  {Perrimon}}]{Gibson2006}%
  \BibitemOpen
  \bibfield  {author} {\bibinfo {author} {\bibfnamefont {M.~C.}\ \bibnamefont
  {Gibson}}, \bibinfo {author} {\bibfnamefont {A.~B.}\ \bibnamefont {Patel}},
  \bibinfo {author} {\bibfnamefont {R.}~\bibnamefont {Nagpal}}, \ and\ \bibinfo
  {author} {\bibfnamefont {N.}~\bibnamefont {Perrimon}},\ }\href {\doibase
  10.1038/nature05014} {\bibfield  {journal} {\bibinfo  {journal} {Nature
  (London)}\ }\textbf {\bibinfo {volume} {442}},\ \bibinfo {pages} {1038}
  (\bibinfo {year} {2006})}\BibitemShut {NoStop}%
\bibitem [{\citenamefont {Lecuit}\ and\ \citenamefont
  {Le~Goff}(2007)}]{Lecuit2007}%
  \BibitemOpen
  \bibfield  {author} {\bibinfo {author} {\bibfnamefont {T.}~\bibnamefont
  {Lecuit}}\ and\ \bibinfo {author} {\bibfnamefont {L.}~\bibnamefont
  {Le~Goff}},\ }\href {\doibase 10.1038/nature06304} {\bibfield  {journal}
  {\bibinfo  {journal} {Nature (London)}\ }\textbf {\bibinfo {volume} {450}},\
  \bibinfo {pages} {189} (\bibinfo {year} {2007})}\BibitemShut {NoStop}%
\bibitem [{\citenamefont {Devany}\ \emph {et~al.}(2021)\citenamefont {Devany},
  \citenamefont {Sussman}, \citenamefont {Yamamoto}, \citenamefont {Manning},\
  and\ \citenamefont {Gardel}}]{Devany2020}%
  \BibitemOpen
  \bibfield  {author} {\bibinfo {author} {\bibfnamefont {J.}~\bibnamefont
  {Devany}}, \bibinfo {author} {\bibfnamefont {D.~M.}\ \bibnamefont {Sussman}},
  \bibinfo {author} {\bibfnamefont {T.}~\bibnamefont {Yamamoto}}, \bibinfo
  {author} {\bibfnamefont {M.~L.}\ \bibnamefont {Manning}}, \ and\ \bibinfo
  {author} {\bibfnamefont {M.~L.}\ \bibnamefont {Gardel}},\ }\href {\doibase
  10.1073/pnas.1917853118} {\bibfield  {journal} {\bibinfo  {journal} {Proc.
  Natl. Acad. Sci. USA}\ }\textbf {\bibinfo {volume} {118}} (\bibinfo {year}
  {2021}),\ 10.1073/pnas.1917853118}\BibitemShut {NoStop}%
\bibitem [{\citenamefont {S{\'a}nchez-Guti{\'e}rrez}\ \emph
  {et~al.}(2016)\citenamefont {S{\'a}nchez-Guti{\'e}rrez}, \citenamefont
  {Tozluoglu}, \citenamefont {Barry}, \citenamefont {Pascual}, \citenamefont
  {Mao},\ and\ \citenamefont {Escudero}}]{Sanchez2016}%
  \BibitemOpen
  \bibfield  {author} {\bibinfo {author} {\bibfnamefont {D.}~\bibnamefont
  {S{\'a}nchez-Guti{\'e}rrez}}, \bibinfo {author} {\bibfnamefont
  {M.}~\bibnamefont {Tozluoglu}}, \bibinfo {author} {\bibfnamefont {J.~D.}\
  \bibnamefont {Barry}}, \bibinfo {author} {\bibfnamefont {A.}~\bibnamefont
  {Pascual}}, \bibinfo {author} {\bibfnamefont {Y.}~\bibnamefont {Mao}}, \ and\
  \bibinfo {author} {\bibfnamefont {L.~M.}\ \bibnamefont {Escudero}},\ }\href
  {\doibase 10.15252/embj.201592374} {\bibfield  {journal} {\bibinfo  {journal}
  {EMBO J.}\ }\textbf {\bibinfo {volume} {35}},\ \bibinfo {pages} {77}
  (\bibinfo {year} {2016})}\BibitemShut {NoStop}%
  \bibitem{Kokic2019}
\BibitemOpen
  \bibfield  {author} {
  \bibinfo {author} {\bibfnamefont {M.}\ \bibnamefont {Kokic}},\
  \bibinfo {author} {\bibfnamefont {A.}\ \bibnamefont {Iannini}},\
  \bibinfo {author} {\bibfnamefont {G.}\ \bibnamefont {Villa-Fombuena}},\
  \bibinfo {author} {\bibfnamefont {F.}\ \bibnamefont {Casares}}\ and\
  \bibinfo {author} {\bibfnamefont {D.}\ \bibnamefont {Iber}},\ }
  \href {\doibase 10.1101/590729}
  {\bibfield  {journal} {\bibinfo {journal} {bioRxiv}\ }
   (\bibinfo {year} {2019})}
   \BibitemShut {NoStop}%
  \bibitem{Dye2021}
\BibitemOpen
  \bibfield  {author} {
  \bibinfo {author} {\bibfnamefont {N.~A.}\ \bibnamefont {Dye}},\
  \bibinfo {author} {\bibfnamefont {M.}\ \bibnamefont {Popovi\'{c}}},\
  \bibinfo {author} {\bibfnamefont {K.}\ \bibnamefont {Venkatesan Iyer}},\
  \bibinfo {author} {\bibfnamefont {J.~F.}\ \bibnamefont {Fuhrmann}},\
  \bibinfo {author} {\bibfnamefont {R.}\ \bibnamefont {Piscitello-G\'{o}mez}},\
  \bibinfo {author} {\bibfnamefont {S.}\ \bibnamefont {Eaton}}\ and\
  \bibinfo {author} {\bibfnamefont {F.}\ \bibnamefont {J\"{u}licher}},\ }
  \href {\doibase 10.7554/eLife.57964}
  {\bibfield  {journal} {\bibinfo {journal} {eLife}\ }
   \textbf {\bibinfo {volume} {10}},\
   \bibinfo {pages} {e57964}
   (\bibinfo {year} {2021})}
   \BibitemShut {NoStop}%
\bibitem{note1}
At each step $j$ of the algorithm, $N\, 2^j$ points are chosen from a uniform
distribution over the domain, which ensures an increasing spatial resolution in
approximation of the centroids as the iterations proceed.%
\bibitem [{\citenamefont {Wada}\ \emph {et~al.}(2011)\citenamefont {Wada},
  \citenamefont {Itoga}, \citenamefont {Okano}, \citenamefont {Yonemura},\ and\
  \citenamefont {Sasaki}}]{Wada2011}%
  \BibitemOpen
  \bibfield  {author} {\bibinfo {author} {\bibfnamefont {K.-I.}\ \bibnamefont
  {Wada}}, \bibinfo {author} {\bibfnamefont {K.}~\bibnamefont {Itoga}},
  \bibinfo {author} {\bibfnamefont {T.}~\bibnamefont {Okano}}, \bibinfo
  {author} {\bibfnamefont {S.}~\bibnamefont {Yonemura}}, \ and\ \bibinfo
  {author} {\bibfnamefont {H.}~\bibnamefont {Sasaki}},\ }\href {\doibase
  10.1242/dev.070987} {\bibfield  {journal} {\bibinfo  {journal} {Development}\
  }\textbf {\bibinfo {volume} {138}},\ \bibinfo {pages} {3907} (\bibinfo {year}
  {2011})}\BibitemShut {NoStop}%
\bibitem [{\citenamefont {Knights}\ \emph {et~al.}(2012)\citenamefont
  {Knights}, \citenamefont {Funnell}, \citenamefont {Crossley},\ and\
  \citenamefont {Pearson}}]{Knights2012}%
  \BibitemOpen
  \bibfield  {author} {\bibinfo {author} {\bibfnamefont {A.~J.}\ \bibnamefont
  {Knights}}, \bibinfo {author} {\bibfnamefont {A.~P.}\ \bibnamefont
  {Funnell}}, \bibinfo {author} {\bibfnamefont {M.}~\bibnamefont {Crossley}}, \
  and\ \bibinfo {author} {\bibfnamefont {R.~C.}\ \bibnamefont {Pearson}},\
  }\href@noop {} {\bibfield  {journal} {\bibinfo  {journal} {Trends Cancer
  Res.}\ }\textbf {\bibinfo {volume} {8}},\ \bibinfo {pages} {61} (\bibinfo
  {year} {2012})}\BibitemShut {NoStop}%
\bibitem [{\citenamefont {Yu}\ and\ \citenamefont {Elble}(2016)}]{Yu2016}%
  \BibitemOpen
  \bibfield  {author} {\bibinfo {author} {\bibfnamefont {Y.}~\bibnamefont
  {Yu}}\ and\ \bibinfo {author} {\bibfnamefont {R.~C.}\ \bibnamefont {Elble}},\
  }\href {\doibase 10.3390/jcm5020026} {\bibfield  {journal} {\bibinfo
  {journal} {J. Clin. Med.}\ }\textbf {\bibinfo {volume} {5}},\ \bibinfo
  {pages} {26} (\bibinfo {year} {2016})}\BibitemShut {NoStop}%
  \bibitem{Classen2005}
\BibitemOpen
  \bibfield  {author} {
  \bibinfo {author} {\bibfnamefont {A.-K.}\ \bibnamefont {Classen}}, \
  \bibinfo {author} {\bibfnamefont {K.~I.}\ \bibnamefont {Anderson}}, \
  \bibinfo {author} {\bibfnamefont {E.}\ \bibnamefont {Marois}}\ and\
  \bibinfo {author} {\bibfnamefont {S.}\ \bibnamefont {Eaton}}, \ }
  \href {\doibase 10.1016/j.devcel.2005.10.016}
  {\bibfield  {journal} {\bibinfo {journal} {Dev. Cell}\ }
   \textbf {\bibinfo {volume} {9}},\
   \bibinfo {pages} {805}
   (\bibinfo {year} {2005})}
   \BibitemShut {NoStop}%
\bibitem{note3}
We note that the anomalous situation where the contact topology of the cells
becoming more heterogeneous over time would result in higher growth rate
is suggestive of the phenomenon of developmental dysplasia in which cells proliferate more
rapidly in a disorderly arrangement.%
\bibitem [{\citenamefont {Janaki}\ \emph {et~al.}(2019)\citenamefont {Janaki},
  \citenamefont {Menon}, \citenamefont {Singh},\ and\ \citenamefont
  {Sinha}}]{Janaki2019}%
  \BibitemOpen
  \bibfield  {author} {\bibinfo {author} {\bibfnamefont {R.}~\bibnamefont
  {Janaki}}, \bibinfo {author} {\bibfnamefont {S.~N.}\ \bibnamefont {Menon}},
  \bibinfo {author} {\bibfnamefont {R.}~\bibnamefont {Singh}}, \ and\ \bibinfo
  {author} {\bibfnamefont {S.}~\bibnamefont {Sinha}},\ }\href {\doibase
  10.1103/PhysRevE.99.052216} {\bibfield  {journal} {\bibinfo  {journal} {Phys.
  Rev. E}\ }\textbf {\bibinfo {volume} {99}},\ \bibinfo {pages} {052216}
  (\bibinfo {year} {2019})}\BibitemShut {NoStop}%
\bibitem [{\citenamefont {Singh}\ and\ \citenamefont
  {Sinha}(2013)}]{Singh2013}%
  \BibitemOpen
  \bibfield  {author} {\bibinfo {author} {\bibfnamefont {R.}~\bibnamefont
  {Singh}}\ and\ \bibinfo {author} {\bibfnamefont {S.}~\bibnamefont {Sinha}},\
  }\href {\doibase 10.1103/PhysRevE.87.012907} {\bibfield  {journal} {\bibinfo
  {journal} {Phys. Rev. E}\ }\textbf {\bibinfo {volume} {87}},\ \bibinfo
  {pages} {012907} (\bibinfo {year} {2013})}\BibitemShut {NoStop}%
\end{thebibliography}

\clearpage
\onecolumngrid

\setcounter{figure}{0}
\renewcommand\thefigure{S\arabic{figure}}
\renewcommand\thetable{S\arabic{table}}

\vspace{1cm}
\begin{center}
\textbf{\large{SUPPLEMENTARY INFORMATION}}\\

\vspace{0.5cm}
\textbf{\large{Disorder in cellular packing can alter proliferation dynamics to regulate growth}}\\

\vspace{0.5cm}
\textbf{Chandrashekar Kuyyamudi, Shakti N. Menon, Fernando Casares and Sitabhra Sinha}
\end{center}
\section*{List of Supplementary Figures}
\begin{enumerate}
\item Fig S1: Increasing homogeneity in the distribution of cell sizes can have differential
outcomes depending on the coupling between the contact-induced signal and the cell cycle
oscillator.
\item Fig S2: Increasing homogeneity in the distribution of number of cellular neighbors can have differential
outcomes depending on the coupling between the contact-induced signal and the cell cycle
oscillator.
\end{enumerate}

\section*{Analysis of the reduced system}
The reduced model obtained after replacing the Hill functions in Eqs. (1-3) in the main text with Heaviside step functions
is described by the following system of coupled piecewise linear equations:
\begin{eqnarray}
  \frac{dA}{dt} &=& [k_1 + k_7 \Theta(A-K)](A_T - A) - k_2 A \, \Theta(C-K), \\
  \frac{dB}{dt} &=& k_3 (B_T - B) - k_4 B \, \Theta(A-K) \,, \\
  \frac{dC}{dt} &=& k_5 (C_T - C) - k_6 C \, \Theta(B-K) - k_8 \frac{S^g}{\Psi^g + S^g} C.
\end{eqnarray}
The behavior of the system can be analyzed by partitioning the entire state space into $(2^3 =) 8$ distinct domains which are defined by the three Heaviside step functions, viz., 
$\Theta(A-K)$, $\Theta(B-K)$ and $\Theta(C-K)$ independently being in any one of
their two possible binary states. This is equivalent to distinguishing the domains by determining
whether each of the variables $A$, $B$ and $C$ have numerical values that are less or 
greater than the constant $K$.
The set of equations that describe the dynamics in each of these $8$ domains are shown 
in Table~\ref{tableS1} below.

\begin{table}[H]
\centering
\begin{tabular}{|l|l|l|l|}\hline
000 & 001  \\ \hline
  &\\
  $\frac{dA}{dt} = k_1(A_T - A)$ & $\frac{dA}{dt} = k_1(A_T - A) - k_2A $     \\
  &\\
  $\frac{dB}{dt} = k_3(B_T - A)$ & $\frac{dB}{dt} = k_3(B_T - A) $      \\
  &\\
  $\frac{dC}{dt} = k_5(C_T - A)- k_8\Phi C$ & $\frac{dC}{dt} = k_5(C_T - A)- k_8\Phi C $         \\
  &\\\hline\hline
  010 & 011 \\ \hline
  $\frac{dA}{dt} = k_1(A_T - A) $       & $\frac{dA}{dt} = k_1(A_T - A) - k_2A $ \\
  &\\
  $\frac{dB}{dt} = k_3(B_T - A) $       & $\frac{dB}{dt} = k_3(B_T - A)$\\
  &\\
  $\frac{dC}{dt} = k_5(C_T - A) - k_6C- k_8\Phi C$ & $\frac{dC}{dt} = k_5(C_T - A) - k_6C - k_8\Phi C$\\
  &\\\hline\hline
  100 & 101 \\ \hline
  &\\
  $\frac{dA}{dt} = k_1(A_T - A) + k_7(A_T - A)$ & $\frac{dA}{dt} = k_1(A_T - A) - k_2A + k_7(A_T - A)$     \\
  &\\
  $\frac{dB}{dt} = k_3(B_T - A) - k_4B$     & $\frac{dB}{dt} = k_3(B_T - A) - k_4B$              \\
  &\\
  $\frac{dC}{dt} = k_5(C_T - A)- k_8\Phi C$            & $\frac{dC}{dt} = k_5(C_T - A)- k_8\Phi C $                   \\
  &\\\hline\hline
  110 & 111 \\ \hline
  $\frac{dA}{dt} = k_1(A_T - A) + k_7(A_T - A)$ & $\frac{dA}{dt} = k_1(A_T - A) - k_2A + k_7(A_T - A) $ \\
  &\\
  $\frac{dB}{dt} = k_3(B_T - A) - k_4B $    & $\frac{dB}{dt} = k_3(B_T - A)- k_4B$ \\
  &\\
  $\frac{dC}{dt} = k_5(C_T - A) - k_6C- k_8\Phi C$     & $\frac{dC}{dt} = k_5(C_T - A) - k_6C- k_8\Phi C$ \\
  &\\ \hline
\end{tabular}
\caption{The system of linear differential equations describing the reduced model in each of the $8$ domains into which the entire state space of the system can be partitioned. The domains are identified by a string of three binary components ($0,1$) that represent the values of
the Heaviside step functions $\Theta(A-K)$, $\Theta(B-K)$ and $\Theta(C-K)$, respectively,
in that domain. Note that the symbol $\Phi$ has been used to represent the 
term ${S^g}/({\Psi^g + S^g})$.}
\label{tableS1}
\end{table}
The focal points $(A^*,B^*,C^*)$ of the system in each of the domains are obtained by solving the
algebraic equations obtained after setting $\frac{dA}{dt} =\frac{dB}{dt} =\frac{dC}{dt} = 0$. 
The focal points thus obtained, as well as the partition in state space that they belong to, 
are listed in Table~\ref{tableS2}. Thus, the global dynamics of the reduced model can be 
represented as a graph of directed flows connecting the $8$ domains [as shown in Fig. 2(a)
in the main text]. The table shows the situation for $k_8 < k^*_8 (=k_5 [(C_T/K) - 1]/\Phi)$
where the system exhibits a cyclic flow involving $6$ of the $8$ domains.
\begin{table}[H]
\centering
\begin{tabular}{|l|c|c|c|c|}\hline
SID & 1 & 2  \\\hline
State & 000 & 001  \\\hline
&&\\
Focal Points & $\left[A_T,B_T,\frac{k_5C_T}{k_5+k_8\Phi}\right]$ & $\left[\frac{k_1A_T}{k_1+k_2},B_T,\frac{k_5C_T}{k_5+k_8\Phi}\right]$   \\
&&\\\hline
Destination & 111 & 011  \\\hline\hline
SID & 3 & 4 \\\hline
State & 010 & 011 \\\hline
&&\\
Focal Points & $\left[A_T,B_T,\frac{k_5C_T}{k_5+k_6+k_8\Phi}\right]$ & $\left[\frac{k_1A_T}{k_1+k_2},B_T,\frac{k_5C_T}{k_5+k_6+k_8\Phi}\right]$ \\
&&\\\hline
Destination & 110 & 010 \\\hline\hline
SID & 5 & 6  \\\hline
State & 100 & 101  \\\hline
&&\\
Focal Points &  $\left[A_T, \frac{k_3B_T}{k_3+k_4},\frac{k_5C_T}{k_5+k_8\Phi} \right]$ & $\left[\frac{(k_1+k_7)A_T}{k_1+k_2+k_7},\frac{k_3B_T}{k_3+k_4},\frac{k_5C_T}{k_5+k_8\Phi} \right]$  \\
&&\\\hline
Destination & 101 & 001 \\\hline\hline
SID & 7 & 8 \\\hline
State & 110 & 111 \\\hline
&&\\
Focal Points & $\left[ A_T, \frac{k_3B_T}{k_3+k_4}, \frac{k_5C_T}{k_5+k_6+k_8\Phi} \right]$ & $\left[\frac{(k_1+k_7)A_T}{k_1+k_2+k_7},\frac{k_3B_T}{k_3+k_4},\frac{k_5C_T}{k_5+k_6+k_8\Phi} \right]$ \\
&&\\\hline
Destination & 100 & 000 \\\hline
\end{tabular}
\caption{The focal points of the system obtained by solving for $A, B, C$ after setting $dA/dt=dB/dt=dC/dt = 0$ are shown for each of the $8$ different domains
into which the entire state space is partitioned. The identification of the
domains by the $3$-bit strings $\{000, \ldots, 111\}$ suggests that the system can be
described in terms of a set of discrete states, and the table indicates which of these states
each of the focal points corresponds to.
Note that the identification of the focal points with the discrete states shown in the table is
for the case $k_8 < k^*_8 (=k_5 [(C_T/K) - 1]/\Phi)$,
when the system exhibits limit cycle oscillations.}
\label{tableS2}
\end{table}
For $k_8 > k^*_8$, the domains $111$ and $101$ flow to $000$, while the states $011$
and $001$ flow to $010$. In turn, both of the states $000$ and $010$ flow to $110$, which
in its turn flows to the state $100$. As the state $100$ maps to itself under the dynamics,
it corresponds to a fixed point attractor into which the system eventually converges to
starting from anywhere in the state space.

The values of the model parameters used
for the simulations whose results are reported in the main text are given in Table~\ref{tableS3}.
\begin{table}[H]
  \centering
  \begin{tabular}{|c|c||c|c|}\hline
  Parameter & Value & Parameter & Value \\\hline
  $k_1$ & 1.0 & $k_2$ & 800 \\
  $k_3$ & 1.0 & $k_4$ & 600 \\
  $k_5$ & 1.0 & $k_6$ & 600 \\
  $k_7$ & 500 & $K$ & 600 \\
  $\Psi$ & 500 & $A_T=B_T=C_T$ & 1000 \\
  $h$ & 4 & $q$ & 4 \\\hline
  \end{tabular}
  \caption{The values for the model parameters used for all simulation results reported.}
\label{tableS3}
\end{table}
\subsection*{Period of oscillations}
As mentioned above (and in the main text) when $k_8< k^*_8$ the system exhibits
oscillations in the concentrations of $A$, $B$ and $C$. The period of oscillation 
can be calculated by noting that it is the sum of the durations spent by the system
in each domain. In other words, we need to add the time intervals between successive 
transitions from one state to another for the cyclic flow between the $6$ discrete states referred
to earlier.
To obtain these intervals, we recall that a transition corresponds to any one of $A$, $B$ or $C$
increasing or decreasing so as to cross the value $K$.
Each transition $j$ is described by a linear equation of the form $\frac{dx_j}{dt} = \alpha_j - \beta_j x$, where $x_j$ represents the concentration of the molecule that
crosses the threshold ($K$). The parameters $\alpha$ and
$\beta$ are functions of the rate constants $k_1, \ldots k_8$, the total concentrations
$A_T, B_T, C_T$, the Hill function parameters $g$, $\psi$ and the strength of the
contact-induced signal $S$ (see Table~\ref{tableS4}). Solving these equations, we obtain
the period as the sum of the time intervals $\tau$ required to switch from one state to another,
with those corresponding
to crossing the threshold from above being given by $\tau (0 \rightarrow 1) = -(1/\beta_j)
\log(1- \{\beta_j K/\alpha_j\})$ and those for crossing the threshold from below being
$\tau (1 \rightarrow 0) = -(1/\beta_j) \log(1- \{[\alpha_j/\beta_j] - K\}/\{[\alpha_j/\beta_j] - T\})$.
\begin{table}[H]
\centering
  \begin{tabular}{|l|c|l|l|}\hline
    $j$ & Transition & $\alpha_j$ & $\beta_j$ \\\hline
    1 & 010 $\rightarrow$ 110  & $k_1 A_T$ & $k_1$ \\
    2 & 110 $\rightarrow$ 100 & $k_3 B_T$ & $k_3 + k_4$ \\
    3 & 100 $\rightarrow$ 101 & $k_5 C_T$ & $k_5 + k_8 \frac{k_8 S^g}{\Psi^g + S^g}$ \\
    4 & 101 $\rightarrow$ 001 & $(k_1+k_7) A_T$ & $k_1+k_2+k_7$ \\
    5 & 001 $\rightarrow$ 011 & $k_3 B_T$ & $k_3$ \\
    6 & 011 $\rightarrow$ 010 & $k_5 C_T$ & $k_5+k_6+k_8 \frac{k_8 S^g}{\Psi^g + S^g}$ \\\hline
  \end{tabular}
\caption{The limit cycle attractor can be broken down into six transitions which can be described by linear
differential equation of the form $\frac{dx_j}{dt} = \alpha_j - \beta_j x$, one for each domain. The parameters $\alpha_j$ and $\beta_j$ required to describe each of the six transitions are shown in the above table.}
\label{tableS4}
\end{table}

\newpage
\section*{Robustness of the proposed mechanism when the signal intensity depends on the cell area}
\begin{figure}[htbp!]
\includegraphics[width=0.8\textwidth]{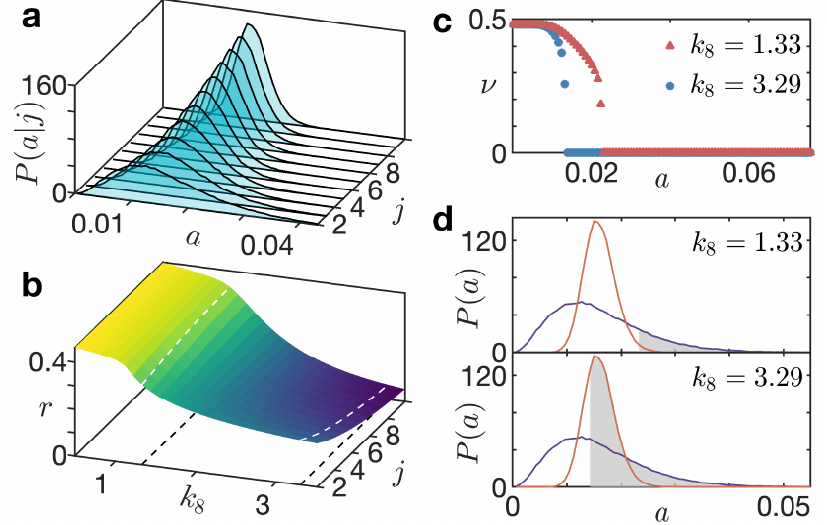}
\caption{\textbf{
Increasing homogeneity in the distribution of cell sizes can have differential
outcomes depending on the coupling between the contact-induced signal and the cell cycle
oscillator.}
(a) The approach to uniformity with successive iterations $j$ is indicated by
the evolution of the distribution of $a$, the areas of the cellular polygons tiling the sheet,
which is seen to become progressively narrower with $j$.
(b) The rate at which the tissue grows by cell division, $r$, given by the mean
of the frequencies of the cellular oscillators, varies with the heterogeneity in
a configuration (determined by the iteration $j$) and the bifurcation parameter $k_8$.
Two contrasting regimes are observed as the polygons become more uniform:
for lower values of $k_8$ the growth rate is seen to increase, while at higher
values, it decreases as the tissue becomes more homogeneous.
(c) The frequency of oscillations $\nu$ of cells having area $a$, shown for the
two regimes,  viz., $k_8 = 1.33$ (triangles) and $k_8 = 3.29$ (circles)
[corresponding to the dashed curves in (b)].
(f) The initial ($j=0$, blue) and final ($j=10$, red) distributions of $a$  corresponding
to the weak (upper panel: $k_8 = 1.33$) and strong coupling (lower panel: $k_8 = 3.29$)
regimes. The shaded region indicates areas above the critical value beyond which the oscillations in cells are arrested.
}
\label{figureS1}
\end{figure}

\newpage
\section*{Robustness of the proposed mechanism when the signal intensity depends on the number of neighbors}
\begin{figure}[htbp!]
\includegraphics[width=0.8\textwidth]{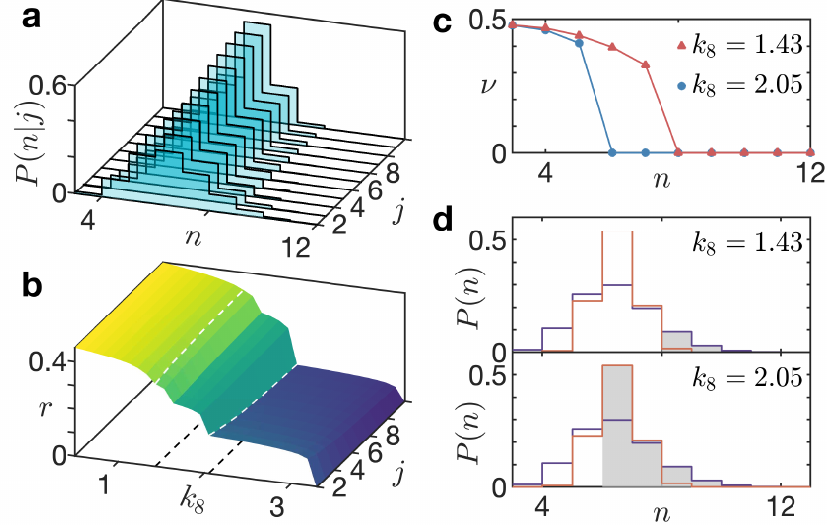}
\caption{\textbf{
Increasing homogeneity in the distribution of number of cellular neighbors can have differential
outcomes depending on the coupling between the contact-induced signal and the cell cycle
oscillator.}
(a) The approach to uniformity with successive iterations $j$ is indicated by
the evolution of the distribution of $n$, the degree (i.e., the number of nearest neighbors
in physical contact with a cell) of the cellular polygons tiling the sheet,
which is seen to become progressively narrower with $j$.
(b) The rate at which the tissue grows by cell division, $r$, given by the mean
of the frequencies of the cellular oscillators, varies with the heterogeneity in
a configuration (determined by the iteration $j$) and the bifurcation parameter $k_8$.
Two contrasting regimes are observed as the polygons become more uniform:
for lower values of $k_8$ the growth rate is seen to increase, while at higher
values, it decreases as the tissue becomes more homogeneous.
(c) The frequency of oscillations $\nu$ of cells having degree $n$, shown for the
two regimes,  viz., $k_8 = 1.43$ (triangles) and $k_8 = 2.05$ (circles)
[corresponding to the dashed curves in (b)].
(f) The initial ($j=0$, blue) and final ($j=10$, red) distributions of $n$  corresponding
to the weak (upper panel: $k_8 = 1.43$) and strong coupling (lower panel: $k_8 = 2.05$)
regimes. The shaded region indicates areas above the critical value beyond which the oscillations in cells are arrested.
}
\label{figureS2}
\end{figure}

\end{document}